\documentclass[twoside]{article}
\pdfoutput=1 
\usepackage{amsmath}
\usepackage{amssymb}
\usepackage{amsthm}
\usepackage{multirow}
\usepackage{graphicx}

\usepackage{placeins}

\newcommand{\indic}[1]{\mathbb{I}(#1)}

\begin{document}



%
%

\title{ Model Selection in Undirected Graphical Models with the Elastic Net }
\author{ 
Mihai Cucuringu%
\thanks{Applied   Mathematics, Princeton University, email: mcucurin@math.princeton.edu}
, Jes\'us Puente%
\thanks{Applied  Mathematics, Princeton University, email: jpuente@math.princeton.edu}
, and David Shue%
\thanks{Computer Science, Princeton University,   email: dshue@princeton.edu}
}

\date{}
\def\ci{\perp\!\!\!\perp}		

\maketitle
 
\begin{abstract}
Structure learning in random fields has attracted considerable attention due to its difficulty and importance in areas such as remote sensing, computational biology, natural language processing, protein networks, and social network analysis. 
We consider the problem of estimating the probabilistic graph structure associated with a Gaussian Markov Random Field (GMRF), the Ising model and the Potts model, by extending previous work on $l_1$ regularized neighborhood estimation to include the elastic net $l_1+l_2$ penalty. Additionally, we show numerical evidence that the edge density plays a role in the graph recovery process. Finally, we introduce a novel method for augmenting neighborhood estimation by  leveraging pair-wise neighborhood union estimates.
\end{abstract}

\section {Introduction}

Edge sparsity in an undirected graphical model (Markov Random Field) encodes conditional independence via graph separation. Essentially, graphical models detangle the global interconnections between the random variables of a joint distribution into localized neighborhoods. Any distribution $P(X)$ consistent with the graphical model must abide by these simplifying constraints. Thus, the graph learning problem is equivalent to a model class selection problem. Let $G=(V,E)$ be an undirected graph on $p=|V(G)|$ vertices and $m=|E(G)|$ edges. Let $X=(X_1, \ldots ,  X_p) \in \mathcal{X}^{p}$ denote a random vector with distribution $P(X)$, where variable $X_i$ is associated to vertex $i \in V(G)$. Graphical model selection attempts to find the simplest graph, often dubbed the {\em concentration graph}, consistent with the underlying distribution.

Recent work in graphical model selection exploits the local structure of the underlying distribution to derive consistent neighborhoods for each random variable. In  terms of graphs, the neighborhood set of a vertex $r$ is $N(r) = \{t \in V(G) \mid (r,t) \in E\}$. More importantly, for undirected graphical models, $N(r)$ is the Markov blanket of $r$, where $X_r$ is rendered conditionally independent of all other variables given $N(r)$: $X_r \ci X_{\setminus (\{r\} \cup N(r))} \mid X_{N(r)}$. 
To estimate the neighborhood conditional probabilities $P(X_r \mid X_{\setminus r})$,  these methods employ pseudo-likelihood measures, specifically $l_1$ regularized regression: the lasso [\ref{tibsh1}]. Compared to other $l_p$ penalty based regularization schemes, the $l_1$ penalty enjoys the dual properties of convexity and sparseness by straddling the boundary between the two domains. 
By treating $X_r$ as the response variable and $X_{\setminus r}$ as the predictors in a generalized linear model, the $l_1$ regularization penalty can recover an appropriately sparse representation of $N(r)$. 
Reconstructing the full edge set of the graph using the estimated neighborhood $\hat{N}(r)$,  allows for two alternate definitions: $ \hat{E}^{\wedge} = \{(a,b):a \in \hat{N}(b) \wedge b \in \hat{N}(a)\} $ which we call AND; or $ \hat{E}^{\vee} = \{(a,b):a \in \hat{N}(b) \vee b \in \hat{N}(a)\}$ which we call OR.

Ravikumar et. al [\ref{ravi1}], [\ref{ravi2}] consider the problem of estimating the graph structure associated with a Gaussian Markov Random Field and the Ising model. Their main result shows that under certain assumptions, the problem of neighborhood selection can be accurately estimated with a sample size of $n=\Omega(d^3 \log p)$ for high dimensional regimes where $d$ is the max degree, and ($p >> n$). Note that the number of samples needed is further improved in [\ref{ravi1}] to $\Omega(d^2 \log p)$ for GMRF model selection. Meinshausen et. al [\ref{graphlasso}] also examine the GMRF case, and provide an asymptotic analysis of consistency under relatively mild conditions along with an alternate $\lambda_1$ penalty.

We build upon this previous work by extending the $l_1$ penalized neighborhood estimation framework to use the elastic net [\ref{elasticnet}] $l_1 + l_2$ penalty and expanding the scope of graphical model recovery to include the multinomial discrete case. While the lasso performs beautifully in many settings, it has its drawbacks. In particular, when $p > n$, the lasso can only select at most $n$ variables. Moreover, for highly correlated covariates, the lasso tends to select a single variable to represent the entire group. By incorporating the $l_2$ penalty term, the elastic net is able to retain the lasso's sparsity while selecting highly correlated variables together. 

Additionally, we introduce a novel scheme for augmenting neighborhood recovery by pooling pair-wise neighborhood union estimates. The idea is to infer the joint neighborhood of a pair of nodes $(i,j)$ (not necessarily adjacent), and obtain the neighborhood of node $i$ by combining all the information given by the $n-1$ pairs of nodes containing node $i$. The frequency with which nodes appear in a specially designed neighbor list for node $i$ gives us a weighted ranking of nodes in terms of their neighbor likelihood. This method can be combined with the usual neighborhood recovery to extract more information from a possibly insufficient set of samples.

\section{Problem formulation}

Undirected graphical models encode the factorization of potential functions over cliques, which in their most basic form, are comprised of 1st and 2nd order interactions: functions that map node and edge values to the real line:
$$P(X) = \frac{1}{Z} \prod_{(s,t) \in E} \phi_{s,t}(X_s, X_t) \prod_{u \in V} \phi_u(X_u), $$
which for max-entropy exponential family distributions, can be written as
$$ P(X) =\frac{1}{Z}  \exp \left( \sum_{(s,t) \in E} \phi_{s,t}(X_s, X_t) + \sum_{u \in V} \phi_u(X_u) \right) $$
Note that $Z$ represents the normalization constant or partition function.

For continuous random variables, the most common exponential family MRF representation is the multivariate Gaussian with sufficient statistics $\{X_s,X_s^2 | s \in V\} \cup \{X_sX_t | (s,t) \in E\}$.
\begin{equation}
P(X) = \frac{1}{Z} \exp \left( \sum_{r \in V} \theta_r X_r  + \frac{1}{2} \sum_{s \in V} \sum_{t \in V} \Theta_{st}X_s X_t \right)
\label{gaussian}
\end{equation}
The $p \times p$ symmetric pairwise parameter matrix $\Theta$, known as the inverse covariance matrix of X denotes the partial correlations between pair of nodes, given the remaining nodes. Every edge $(s,t) \in E$ will have a non-zero entry in $\Theta$ and each row $s$ of $\Theta$ specifies the graph neighborhood $N(s)$. Conversely, the sparsity of $\Theta$ reveals the conditional independencies of the graph where $\Theta_{st} = 0,  \forall (s,t) \notin E$.  Conditional neighborhood expectations can be represented by a linear model: $\mathbb{E}(X_s | X_{\setminus s}) = \sum_{t \in N(s)} \theta_{st}X_t$.

In the binary case, the MRF distribution can be described using an Ising model where $ X_s \in \{ -1,1\}, \forall s \in V$, and $\phi_{st}(X_s, X_t) = \theta_{st} X_s X_t$. The full probability distribution takes the following form, which omits first order terms:
\begin{equation}	
P(X) = \frac{1}{Z} \exp \left( \sum_{(s,t) \in E} \theta_{st} X_s X_t \right)
\label{isingdist}
\end{equation}	
The conditional neighborhood probability $P(X_s | X_{\setminus s})$ is defined as:
\begin{equation}
P(X_s | X_{\setminus s}) = \frac{\exp (2X_s \sum_{(s,t) \in E} \theta_{st}X_t)} {\exp (2X_s \sum_{(s,t) \in E} \theta_{st}X_t) + 1}
\label{isingneighborhood}
\end{equation}
Taking the Hessian of the local conditional probability gives the Fisher information matrix for $X_s$, 
Much like partial correlations in the Gaussian concentration matrix, zero entries in the Fisher information matrix indicate conditional independence.

Extending the discrete parameterization to variables with $k>2$ states, requires an expansion in terms where the edge potential functions $\phi$ now describe a set of parameterized indicator variables $\mathbb{I}(X_s = l,X_t = m)$ representing the $k^2$ possible value pairs between $X_s$ and $X_t$.
$$ P(X) =\frac{1}{Z} \exp \left( \sum_{s \in V} \sum_{i = 1}^{k} \theta_{s:i} \mathbb{I}(X_s = i) \right.  \left.  +  \sum_{(s,t) \in E} \sum_{l = 1}^{k} \sum_{m = 1}^{k}    \theta_{st:lm} \mathbb{I}(X_s = l, X_t = m) \right)
 \label{discrete}$$
As described in [\ref{expfamily}], this particular representation is over complete since the indicator functions satisfy a variety of linear relationships  $\sum_{i = 1}^{k} \mathbb{I}_{s}(X_s = i) = 1$. However, despite the lack of a guaranteed unique solution, the factorization can still satisfy the desired neighborhood recovery criterion. 
A simplified variant of the general discrete parameterization is the Potts model where each $\phi$ is defined by two indicator functions denoting node agreement and disagreement for arbitrary $k > 2$. We observe that in the Ising model, the form of $\phi$ may be be recast as $\phi_{st}(X_s,X_t)=\theta_{st}\indic{X_s=X_t}-\theta_{st}\indic{X_s\neq X_t}$. 
$$
P(X) =\frac{1}{Z} \exp \left( \sum_{s \in V} \sum_{i = 1}^{k} \theta_{s:i} \mathbb{I}(X_s = i) \right. +  \left. \sum_{(s,t) \in E}  \theta_{st}\indic{X_s=X_t}-\theta_{st}\indic{X_s\neq X_t} \right)
\label{potts}
$$
Note that the Potts model only requires a single parameter and generalizes the Ising model to $k$ states.



%

To extend neighborhood estimation from the binary Ising model case to a discrete parameterization, we note that the neighborhood conditional probability takes the form 
\begin{equation}
P(X_s = d \mid X_{\setminus s} )= \frac { \exp \{  \theta_{s:d} +  \sum_{(s,t) \in E}  \sum_{m = 1}^{k} \theta_{st:dm} \mathbb{I}(X_s = d, X_t = m) \}} { \sum_{l = 1}^{k} \exp \{ \theta_{s:l} +  \sum_{(s,t) \in E} \sum_{m = 1}^{k}    \theta_{st:lm} \mathbb{I}(X_s = l, X_t = m) \} }
\end{equation}
which is equivalent, after a variable transformation from a discrete feature space to indicators, to the classical multinomial logistic regression equation:
\begin{equation} 
P(X_s = d \mid X_{\setminus s}) = \frac { \exp \{  \theta_{d}^{T}X_{N(s),d}\}} { \sum_{l = 1}^{k} \exp \{ \theta_{l}^{T} X_{N(s),l} \} }
\end{equation}
where $X_{N(s), l} = \{\mathbb{I}(X_s = l, X_t = m) \mid (s,t) \in E\}$ with an additional singleton indicator variable $\mathbb{I}(X_s = l)$ always set to 1. 
With the conditional probability equations in hand, we can approach the problem of neighborhood estimation as a generalized linear regression.
Building on previous model selection work using the $l_1$ penalty, we extend the approach, to use the combined $l_1+ l_2$ penalty approach of the elastic net [\ref{elasticnet}], which for the basic linear model takes the form:
$$ L(\lambda_1, \lambda_2,\theta) = ||x_s -X_{\setminus s}\theta||_{2}^2 + \lambda_1 ||\theta||_{1} + \lambda_2 ||\theta||^{2}_{2} $$
The elastic net performance surpasses the $l_1$ penalty under noisy conditions and where groups of highly correlated variables exist in the graph. However, as noted by Bunea [\ref{bunea}], the additional $l_2$ smoothing penalty should be small relative to the $l_1$ term to preserve sparsity. Many authors have extended the elastic net penalty to additional regression models, covering a broad swath of the generalized linear realm. For the linear Gaussian case, we use the original elastic net package of Zhou and Hastie [\ref{elasticnet}].  For binary and multinomial regression we rely on the glmnet library of Friedman et. al [\ref{fried1}]. 

\section{Experimental Evaluation}


To evaluate the elastic net for Gaussian MRF model selection, we generate the distribution inverse covariance matrix $ \Theta = \Sigma ^{-1}$ in the following way. We set $ \Theta_{ij} = 0.5$ whenever $(ij) \in E$, and then perturb the diagonal of the matrix $\Theta_{ii} = \tau$, with $\tau$ large enough to force all eigenvalues of $\Theta$ to pe positive. We experimentally choose $\tau$, starting from $1$ and increasing it in increments of $0.1$ until we get a value that makes $\Theta$ positive definite.  


In the case of the binary and discrete models,  we require a more complicated procedure based on MCMC sampling. However, given the size of our graphs, the direct Gibbs sampling approach proved to be computationally expensive because of its long mixing times and slow mode exploration when the temperature (the $\theta$'s in our case) is low.

To overcome this difficulty, we turn to the Swendsen-Wang algorithm. This method generates an augmented graph $\tilde{G}=(\tilde{V},\tilde{E})$, where $\tilde{V}=V \cup E$ and $\tilde{E}$ contains $(v,e)$ iff $v\in V$ and $e \in E$ are incident. Given this formulation, $\tilde{G}$ is bipartite between the $V$ nodes and $E$ nodes. Thus in the joint distribution of $\tilde{G}$, the Markov blanket of $E$ will only consist of elements in $V$ and vice versa. The random variables assigned to $E$ can only take the values 0 and 1.

We  define the conditional probabilities of $V$ and $E$ as:
\begin{itemize}
\item $P_{ \tilde{G} }(e=1|V)$ is given by considering the nodes $s,t\in G$ s.t. $e$ is incident with $s$ and $t$. $P_{\tilde{G}}(e=1|V)=1-2e^{-2J}$ if $x_s=x_t$ and 0 otherwise.
\item $P_{ \tilde{G} }(V|E)$ is such that all nodes in the same component (in the graph when we consider only the edges $e$ s.t. $e=1$) have the same value, and each component takes each of the $k$ possible values with equal probability.
\end{itemize}
Essentially, the algorithm generates MCMC samples by alternately updating the values of $V$ and $E$ using Gibbs sampling. Although the augmentation substantially increases the number of vertices, the algorithm creates a Markov chain that explores the space of outcomes much more rapidly. For the details of the Swendsen-Wang algorithm, we refer the reader to [\ref{swendsenwang}] and [\ref{mackay}]. 

By introducing an $l_2$ penalty term to the regression model, the maximization problem in the elastic net setup becomes 
\begin{equation} \label{enetGMRFeq}
\hat{\theta}^{s,\lambda_1,\lambda_2 } = \mathrm{arg}\min_{\theta: \theta_{s}=0}  \| X_s - X\theta\|_2^2 + \lambda_1 \|\theta\|_1 + \lambda_2 \|\theta\|_2^2 
\end{equation}
with $\lambda_2 = c \sqrt(\frac{\log p}{n})$. We experimented with several values of $\lambda_2$ for different number of samples, and observed the type I and type II probability errors.

\begin{figure}[!ht]
\centering
\includegraphics[width=0.4 \textwidth]{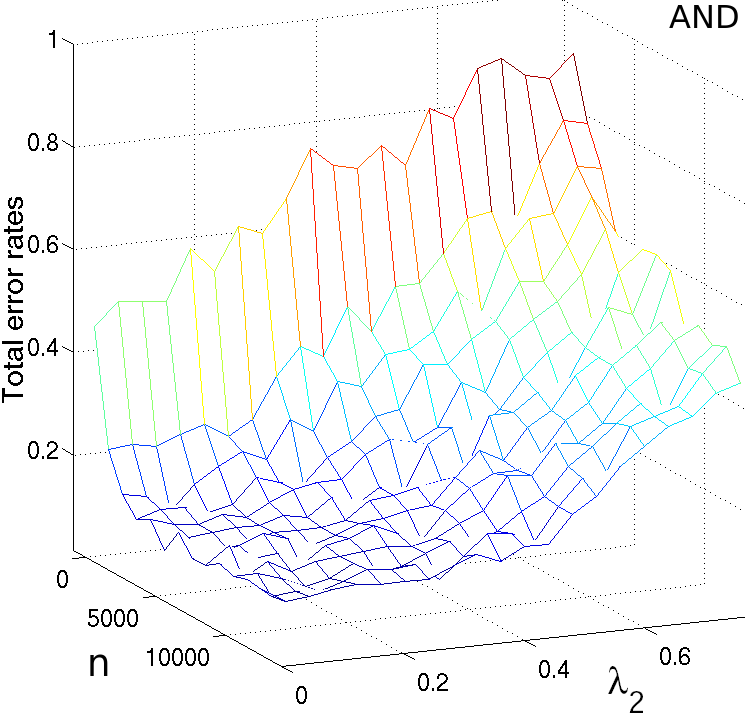}
\includegraphics[width=0.4 \textwidth]{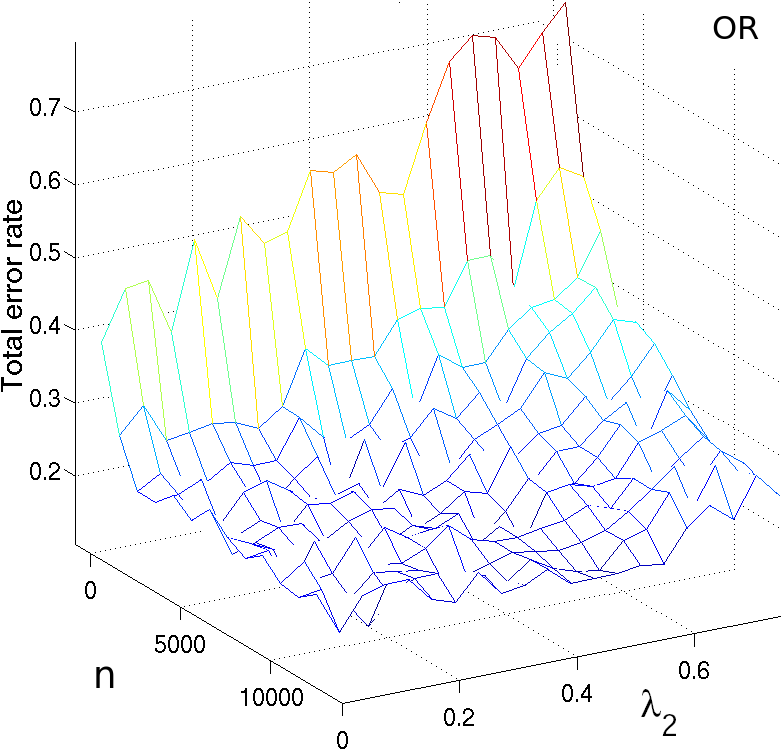}
\caption{ Total error rates as a function of $\lambda_2$ and number of samples $n$, in the AND neighborhood selection (left) and the OR neighborhood selection (right).}
\label{surfaceandor}
\end{figure}

\begin{figure}[!ht]
\centering
\includegraphics[width=0.8 \textwidth]{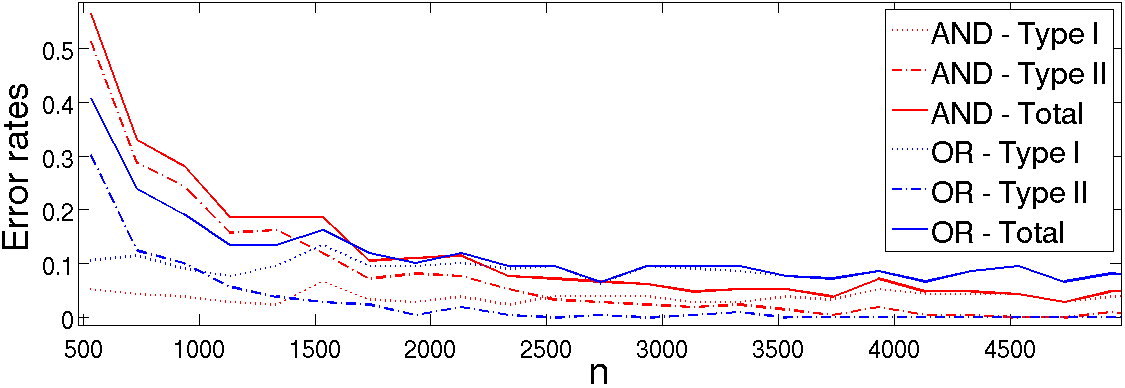}
\caption{ Error rates for a random graph, with $p=40$, $\lambda_2 = \sqrt{\frac{log(p)}{n}}$, and $n \leq 5d^2log(p)$.}
\label{p40errvsn}
\end{figure}

In Figure \ref{surfaceandor} we show 3D plots of the total error rates as a function of the number of samples and the $\lambda_2$ parameter, for the AND and OR neighborhood estimation. Several observations can be made from these plots. First note that larger values of $\lambda_2$ performs worse than smaller values, no matter what the sample size is. The best recovery rates are achieved when $\lambda_2$ is very small. Also note that the AND neighborhood selection perform much worse than its OR counterpart when $\lambda_2$ is large.
The figure \ref{p40errvsn} plots the error rates versus the sample size, for a fixed $\lambda_2 = c \sqrt(\frac{\log p}{n})$. Note that the chosen graph has $40$ vertices and is of maximum degree $d=25$, and it cannot be recovered without errors even when the number of samples scales as $5d^2 log(p)$.

The first family of graphs we tested were the star graphs and the more general clique of star graphs. We denote by $Star(a,b)$ the clique-star graph obtained from $a$ copies of a star graph\footnote{A star graph consists of one node of degree $q$ connected to the remaining $q$ vertices of degree $1$} by connecting all star centers among them, or in other words we add $d$ different neighbors to each vertex in a clique of size $a$. $Star(1,b)$ is just the standard star graph. Note that the maximum degree in $Star(a,b)$ is $d = a+b-1$ and the total number of vertices is $p=a(b+1)$.

For a graph $G$ we let $\rho(G)$ denote the edge density of the graph, i.e. $\rho = \frac{2|E|}{n(n-1)}$. The reason we introduce this parameter in our simulations is to observe the impact of edge density  on the recovery rates when the maximum degree and the number of samples are fixed. As our simulations show, recovering the graph structure is significantly harder when the graph has a higher density but the same fixed maximum degree $d$. To test this, we generate a star graph with maximum degree $d$ and edge density $\rho_1$, and then start adding edges among the lower degree neighbors to obtain a new graph with new edge density $\rho_2 > \rho_1$. Note that this edge density dependence can be equivalently formulated in terms of the average degree $\bar{d}$ of a graph by the formula $\bar{d} = (p-1) \rho$. The error rates are averaged over $20$ runs for a fixed graph $G$ on different samples of size $n$. Additionally, $\lambda_2$ was chosen by discretizing the interval $[0, \sqrt{\frac{\log p}{n}}]$ into $15$ equally sized subintervals.

\begin{figure}[!ht] \centering
\includegraphics[width=0.8\textwidth]{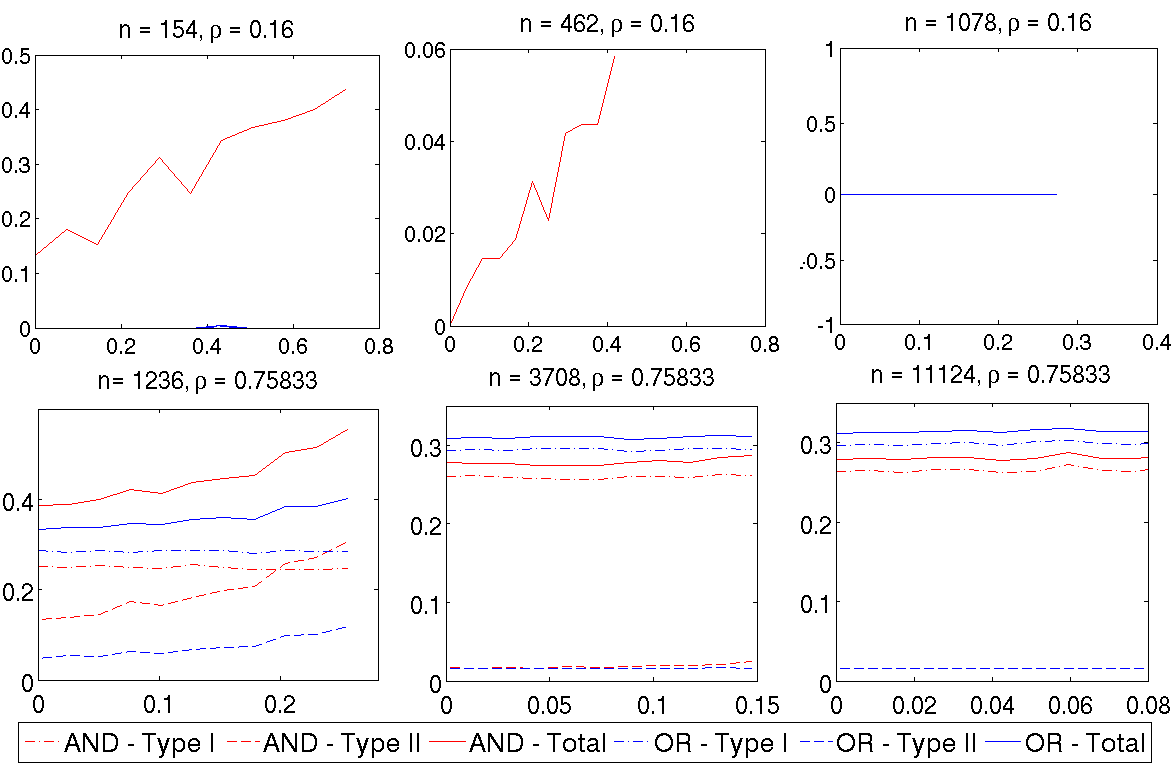}
\caption{ Error recovery rates ($y$-axis) versus the $\lambda_2$ parameter ($x$-axis) for the graphs $G_1 = Star(1, 24)$ with $\rho_1=0.16$ (top) and $G_2$ with $\rho_2=0.76$ (bottom).}
\label{star2rhos}
\end{figure}

\begin{figure}[!ht]
\centering
\includegraphics[width=0.8 \textwidth]{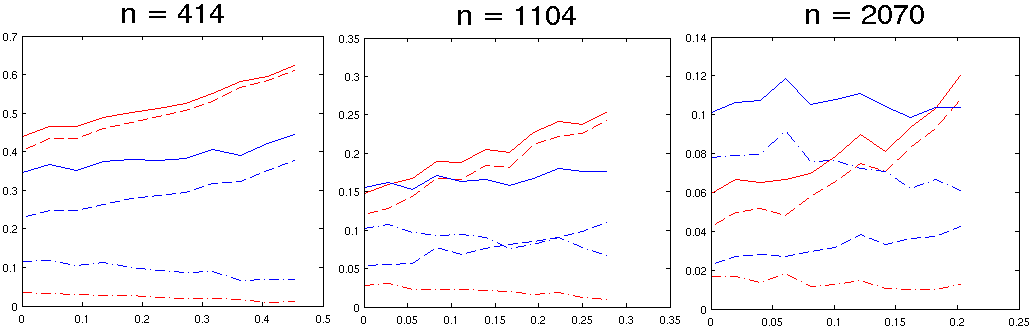}
\caption{ Error recovery rates ($y$-axis) versus the $\lambda_2$ parameter ($x$-axis) for the graph $H_2$ with $\rho_1 = 0.30$.}
\label{wstar2rhos}
\end{figure}

Figure \ref{star2rhos} plots the error recovery rates for $G_1 = Star(1,24)$ with $\rho_1 = 0.16$ (top) and for the graph obtained by adding edges to $G_2 = Star(1,24)$ until $\rho_2 = 0.76$(bottom), with both graphs having the same maximum degree $d=24$. Note that for these graphs $d^2 log(p) \approx 1800$ and as seen in the top plot of Figure \ref{star2rhos}, a bit over $1000$ samples are enough to bring the error rates to zero. However, the bottom shows that even with $10$ times more samples, we can only recover $G_2$ with a $0.30$ error rate. We repeat the above experiment for the clique-star graph $H_1 = Star(6,4)$ with $p=30$ and edge density $\rho_1=0.18$, and the graph $H_2$ obtained by adding edges to $H_1$ while keeping the maximum degree $d=9$ unchanged. In this case, $d^2 log(p)=275$ and we successfully recover $H_1$ only when the sample size exceeds $1400$, due to higher edge density (plot omitted). Figure \ref{wstar2rhos} shows the error rates when we increase the edge density to $\rho_1 = 0.3$, which emphasizes the increase in sample size required for graph recovery.
Note that in both simulations the $\lambda_2$ penalty was rarely of any help, and in most cases $\lambda_2 = 0$ achieved the best error rates.

\begin{figure}[!ht]  
 \centering
 \includegraphics[width=0.8 \textwidth]{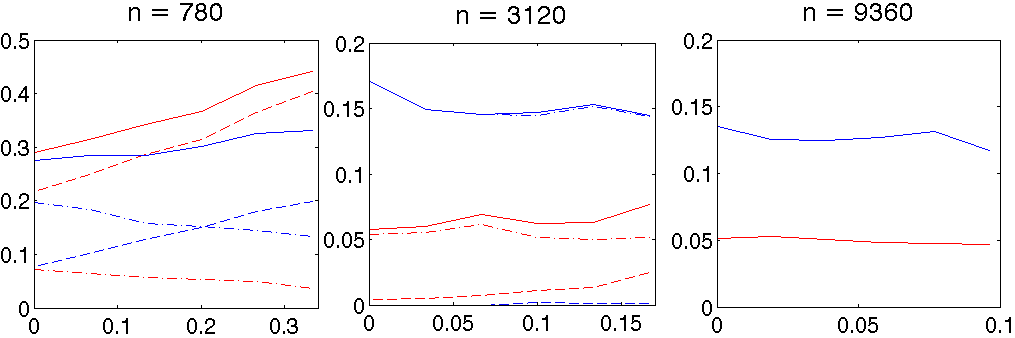}
\caption{ Error rates for the community graph $Com(4,8,0.8,0.15)$ with $p = 32$, $d = 15$, and $\rho = 0.28$}.
\label{com32}
\end{figure}

A second type of graph we considered was the community graph, denoted by $Com(s,t,\beta_{in},\beta_{out})$, which consists of $s$ groups of highly connected nodes, where each group has size $t$, so $p=st$. Two vertices within the same group (or community) are connected with probability $\beta_{in}$, while nodes that belong to two different communities share an edge with a smaller probability $\beta_{out}$. These community structures are a common feature of complex networks, and have the property that nodes within a group are much more connected to each other than to the rest of the network (for $\beta_{in} > \beta_{out}$). In application, these communities may represent groups of related individuals in social networks, topically related web pages or biochemical pathways, and thus their identification is of central importance. To completely understand the modular structure of such graphs, one should be able to both detect overlapping communities and make meaningful statements about their hierarchies [\ref{ohcom}]. Figure \ref{com32} plots the error rates in the recovery of a $Com(4,8,0.8,0.15)$ graph with $p = 32$, $d = 15$, and $\rho=0.28$. $d^2 log(p) = 780$, however again even with 9000 samples, the error rate is still over 10 percent. When few samples are available, $\lambda_2 = 0$ achieves the best error rates, but as we increase the number of samples we notice that the elastic net method with $\lambda_2 >0 $ performs slightly better than when $\lambda_2=0$. While this improvement is not significant, it hints that the additional $l_2$ regularization may produce better results in some cases. 

\section{Discrete MRF recovery}

\begin{figure}[!ht]  
 \centering
\includegraphics[width=0.8 \textwidth]{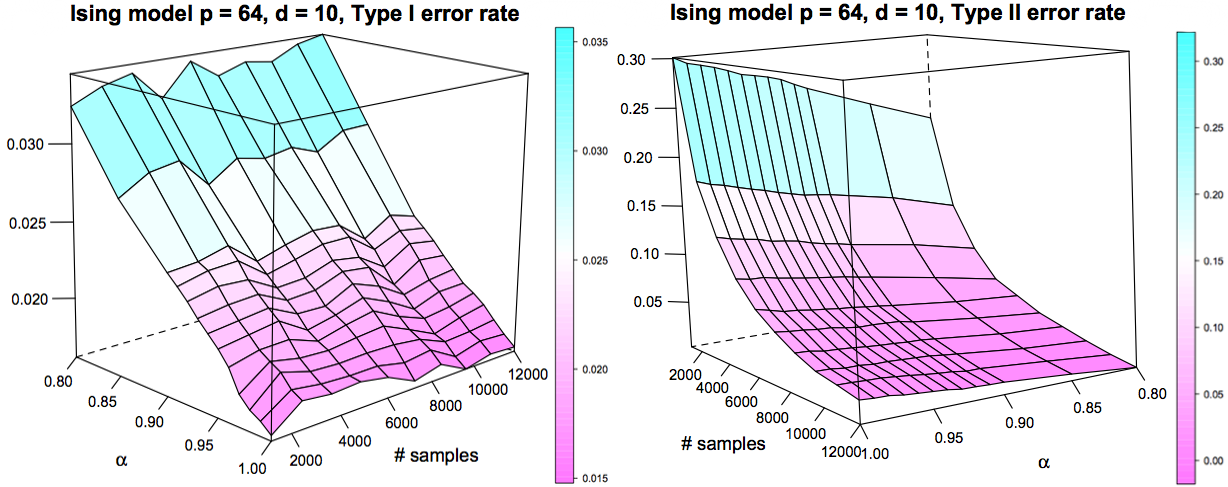}
\caption{Type I and II error rate for an Ising model graph with 64 nodes and max degree 10. While $\alpha \approx 1$ minimizes FPR across sample sizes, the conservative nature of the $l_1$ penalty induces false negatives even for large sample sizes. Choosing an $\alpha < 1$ increases the likelihood of recovering additional edges}
\label{ising64}
\end{figure}

\begin{figure}[!ht]  
 \centering
 \includegraphics[width=0.8 \textwidth]{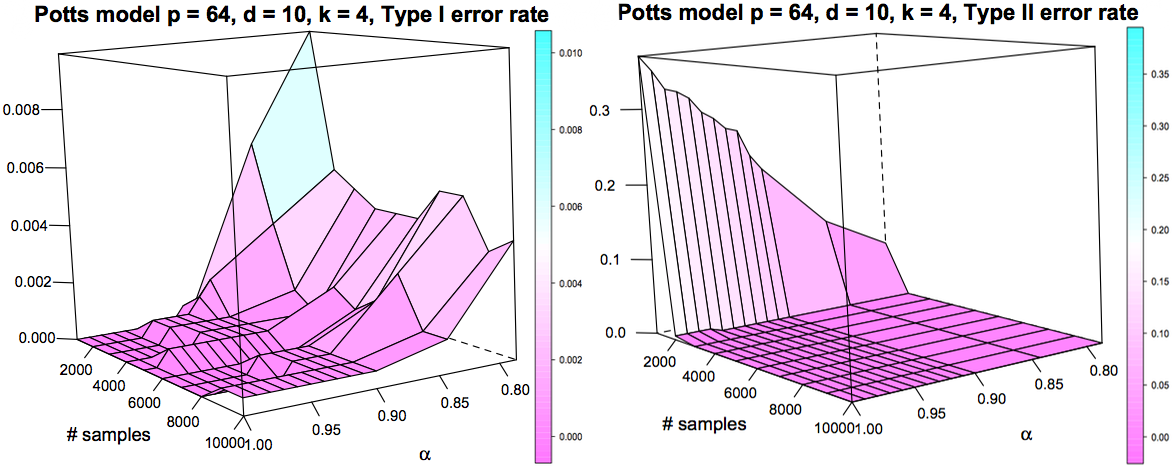}
\caption{Type I and II error rate for a Potts model graph with 64 nodes, max degree 10, and 4 states. When $\alpha = 1$ the FPR is minimized over all sample sizes. However, again, the $l_1$ penalty induces false negatives for small sample sizes}
\label{potts64}
\end{figure}

Figures \ref{ising64} and \ref{potts64} depict the performance of the elastic net neighborhood estimator over a range of discrete MRF graphs. The graphs evaluated for the Ising and Potts model are random graphs with bounded maximum degree. All experiments were run over a sample range covering the $d^3 \log p$ edge recovery threshold and $\alpha$ values ranging from 0.5 to 1, where $\alpha = \frac{\lambda_1}{\lambda_1 + \lambda_2}$. Results from multiple trials were averaged for the Ising model. Due to the computational load of the \textit{glmnet} multinomial regression for large data sizes, only single run results are shown for the Potts model. Smaller $\alpha$ values are omitted from the plot in order to limit the scale and improve clarity.  Unless otherwise noted, the plots represent AND neighborhood unions, with OR neighborhood unions showing similar performance. 

As seen in the plots, the elastic net neighborhood estimator recovers the underlying graph with high probability under corresponding $\Omega(d^3 \log p)$ sample sizes, validating our formulation of the discrete model neighborhood estimation as a multinomial logistic regression. Similar to the Gaussian case, the effect of the $l_2$ penalty $\lambda_2$, (while $\lambda_1$ is set to $\sqrt{\frac{\log(p)}{n}}$) tends to benefit neighborhood recovery mostly at small sample values and when $\alpha$ is close to 1. Oversized $l_2$ penalties introduce an inordinate number of noise edges, but small $l_2$ penalties reduce the chance of missing edges with weak correlation which the $l_1$ penalty rejects. When the $l_2$ penalty is non-zero, the minimization function is strictly convex and allows the estimator to select additional nodes that exhibit highly correlated behavior by effectively averaging their contribution. 

\begin{figure}[!ht]  
 \centering
 \includegraphics[width=0.8 \textwidth]{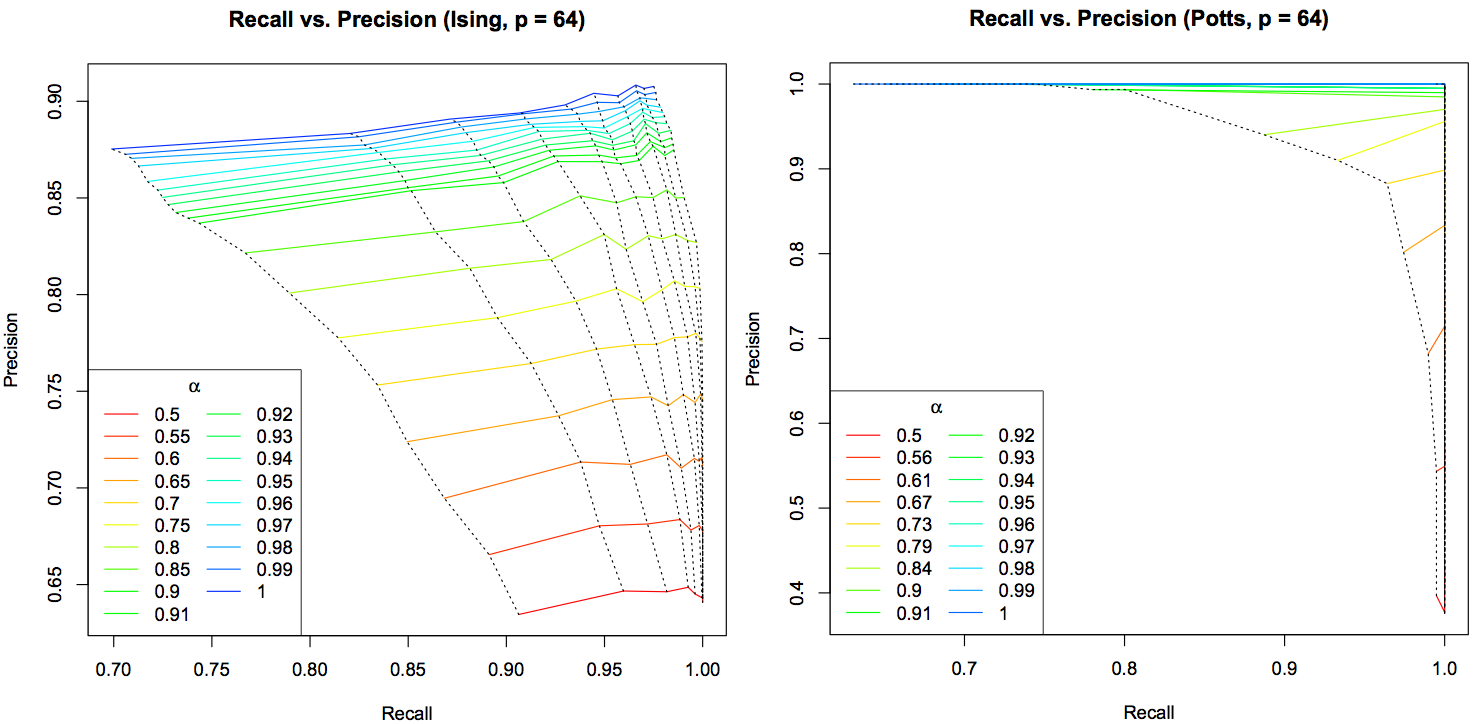}
\caption{Recall-Precision curve for Ising and Potts model with 64 nodes and max degree 10. Horizontal curves correspond to $\alpha$ values varying over sample sizes represented by the vertical dashed black curves increasing to the right. The left most vertical curve corresponds to sample size 1200. Although the $\alpha = 1$ pure $l_1$ penalty dominates precision performance, choosing a smaller $\alpha$ permits a trade-off in precision for recall that is particularly beneficial for small sample sizes}
\label{isingpotts64RP}
\end{figure}

The $\lambda_2$ parameter provides, in essence, a trade-off between precision and recall, as it can be seen in Figure \ref{isingpotts64RP}, especially when the number of samples is small, which is the case in many high dimensional $p \gg n$ applications.  While the $\alpha = 1$ curve consistently provides the highest graph recovery precision over all sample sizes, the actual number of recovered edges may be extremely limited due to the sparsity constraint. For the Ising model graph depicted in the figure, the smallest sample size 1200 with $\alpha = 1$ gives a precision of 0.88 with a recall of 0.7. By introducing a $\lambda_2$ term with $\alpha = 0.8$, the precision drops to 0.8 but the recall improves to 0.79, which is better than a 1 to 1 trade-off. As expected, with large sample sizes, the benefit of the $\lambda_2$ parameter diminishes as shown by the nearly vertical slope of the large sample size curves. Similarly, the Potts model recall-precision plot also displays this trend, albeit in a compressed fashion since the neighborhood estimator is able to recover the graph at a smaller sample size, rendering the larger sample size curves uninformative. From these results we can say that the additional presence of an $\l_2$ penalty may yield substantial benefits for $p \gg n$ situations where the goal is to extract relevant correlation information from small sample sizes.

\section{Neighbors of pair of vertices}
The technique introduced in this section is useful in neighborhood reconstruction especially in the case of regular graphs, graphs with a small gap between known maximum and minimum degree, or when we would like to obtain a likelihood ranking of the $p-1$ possible neighbors of a fixed node $i$.
The idea is to infer neighborhoods not only for one vertex at a time, but for pairs of vertices, which may or may not be adjacent. We denote by $\hat{\mathcal{N}}^{1}(i)$ the estimation of the neighborhood of node $i$, as given by the optimization in equation (\ref{enetGMRFeq}).

We denote by $\mathcal{N}_{ij}$ the set of neighbors of nodes $i$ and $j$, i.e. $\mathcal{N}_{ij} = \{ v \in V(G) \backslash \{i,j\} : (i,v) \in E \text{ or } (j,v) \in E \}$, in other words $\mathcal{N}_{ij}$ is the union of the neighborhoods of nodes $i$ and $j$, minus the edge $(i,j)$ if it exists.
We now define the following optimization problem similar to the one in equation (\ref{enetGMRFeq})
 
\begin{equation}
(\hat{\theta}^{I} , \hat{\theta}^{J})^{\lambda_1,\lambda_2}:=  \mathrm{arg}\min_{\theta^{I}, \theta^{J}: \theta^{I}_{i}=0, \theta^{J}_{j}=0 }    \| X_i - X\theta^{I} +  X_j - X \theta^{J} \|_2^2 + \lambda_1 \|\theta^{I}\|_1  $$$$+ \lambda_2 \|\theta^{I}\|_2^2  +  \lambda_1 \|\theta^{J}\|_1 + \lambda_2 \|\theta^{J}\|_2^2 
\label{tij}
\end{equation}
After grouping the terms of $\lambda_1$ and $\lambda_2$ and approximating $ \|\theta^{I}\|_1 + \|\theta^{J}\|_1 $ by $\|\theta^{I} + \theta^{J}\|_1 $, and $ \|\theta^{I}\|_1 + \|\theta^{J}\|_2^2 $ by $\|\theta^{I} + \theta^{J}\|_2^2 $,  we approximate (\ref{tij}) by 
\begin{equation}
    \mathrm{arg}\min_{\theta, \theta_{i}=0, \theta_{j}=0 }  \| (X_i + X_j) - X \theta  \|_2^2  + \lambda_1 \|\theta\|_1 + \lambda_2 \|\theta\|_2^2   =:   \hat{\theta}^{I,J, \lambda_1,\lambda_2} 
\end{equation}
where in the last step we make the change of variable $\theta = \theta^{I} + \theta^{J}$ and we denote by $\hat{\theta}^{I,J, \lambda_1,\lambda_2}$ the regression coefficients of the sum of variables $I$ and $J$ against the remaining variable. We now define the estimated neighborhood of a pair of vertices $(i,j)$, not necessarily adjacent, to be $\hat{\mathcal{N}}_{ij} = \{ v \in V(G) \backslash \{i,j\} :   \hat{\theta}^{ij,\lambda_1,\lambda_2 }(v) \neq 0 \}$, in other words $\hat{\mathcal{N}}_{ij}$ is an estimate of $\mathcal{N}_{ij}$. Note that for a vertex $ v \in \hat{\mathcal{N}}_{ij}$ it may be the case that $v$ is adjacent to either $i$ or $j$ or perhaps both.   

For a fixed node $i$, we obtain its neighborhood in the following way. We let $\mathcal{L}_{i}$ denote the list obtained by concatenating the pair neighborhoods of node $i$
				$$ \mathcal{L}_{i} = \bigsqcup_{j \neq i} \mathcal{N}_{ij} $$
where $\bigsqcup$ denotes union with repetitions. We denote by $\hat{\mathcal{L}}_{i}$ the concatenation of the estimated pair neighborhoods, i.e. $ \hat{\mathcal{L}}_{i} = \bigsqcup_{j \neq i} \hat{\mathcal{N}}_{ij} $. 
Note that $\mathcal{L}_{i}$ includes all nodes that are neighbors of $i$, each appearing with multiplicity $p-2$. If $j$ is a neighbor of $i$, then $j \in \mathcal{N}_{ik}$ for all $k \neq i,j$, i.e. $p-2$ times. In the absence of errors $ \hat{\mathcal{N}}_{ij} = \mathcal{N}_{ij}$,  $ \hat{\mathcal{L}}_{i}= \mathcal{L}_{i}$, and with the exception described in the next paragraph, we can correctly recover the neighborhood of node $i$ by picking the most frequent elements from $ \hat{\mathcal{L}}_{i} $, i.e. all nodes which appear in the list exactly $p-2$ times. In the case of errors, we obtain an estimate for the neighborhood of node $i$ by selecting the most frequent elements in $ \hat{\mathcal{L}}_{i}$. Also, if there are no errors, $\mathcal{L}_{i}$ contains all other nodes $j \neq i$ of $G$ at least once. This is obvious if $j \sim i$, as $j$ appears $p-2$ times in $\mathcal{L}_{i} $ as explained above. If $j \nsim i$, then pick $k$ a neighbor of $j$ ($k$ exists since we assumed $G$ is connected) and it must be that $ j \in \mathcal{N}_{ik} \subseteq \mathcal{L}_{i}$.

Note that a non-neighbor node $j$ of $i$ can appear $n-2$ times in $ \mathcal{L}_{i}$ if $j$ is connected to all nodes in $ V(G) \backslash i$, in which case we (incorrectly) add $j$ to the neighborhood of node $i$. Similarly, if $i$ is connected to all nodes in $ V(G) \backslash j$, then $i$ appears in $\mathcal{L}_{j}$ with multiplicity $p-2$ and we (incorrectly) mark $i$ as being in the neighborhood of node $j$. In other words, $i$ and $j$ appear in each other's neighborhood lists, thus rendering our approach incorrect, whenever $i$ and $j$ are both connected to all other $p-2$ vertices in the graph. However, for random graphs this scenario occurs with a very low probability.

\begin{figure}[h] 
 \centering
 \includegraphics[width=.8 \textwidth]{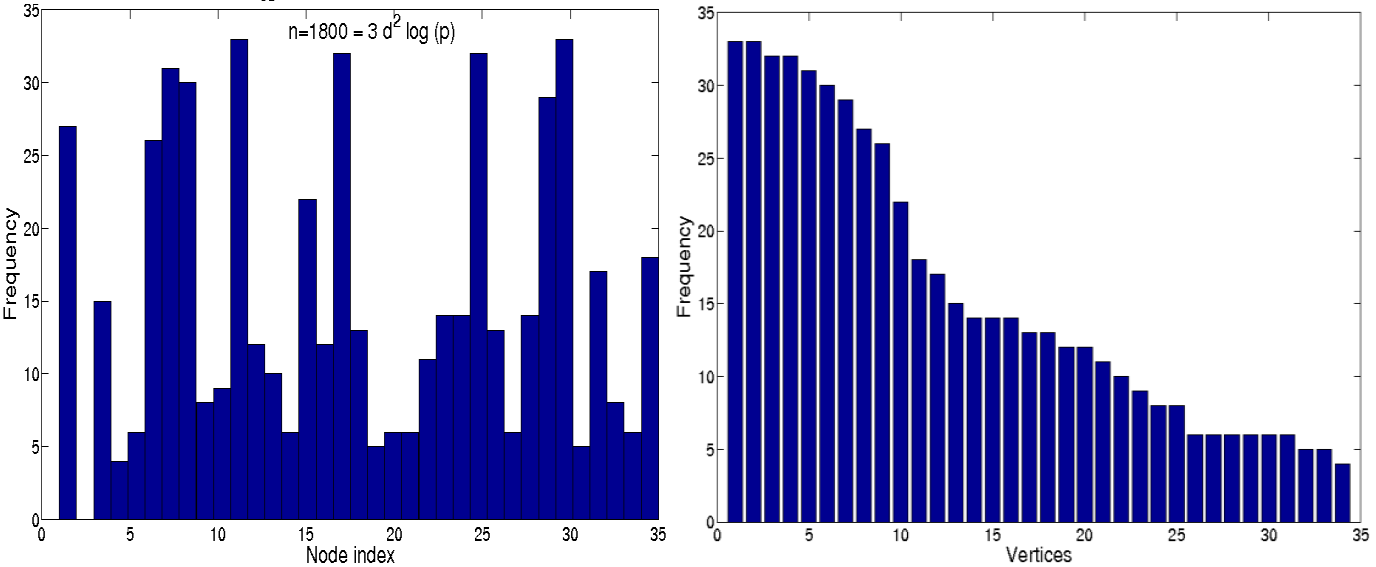}
\caption{ $\mathcal{N}_{2} = \{  1,  6,   7,   8,  11,  17,  25,  29, 30 \}$, 
$ \hat {\mathcal{N}_{2}} = \{ 1, 6, 7,  8,  11,  15, 17,  25,  29,  30 \} $,
$\hat{\mathcal{L}}_{2} =  \{ $  \textbf{30(33), 11(33), 25(32), 17(32), 7(31), 8(30), 29(29), 1(27), 6(26)}, 15(22), 35(18), 32(17),  $ \ldots  \}$ where boldface indicates true neighbors and numbers in parentheses denote frequencies in $\hat{\mathcal{L}}_{2}$}
 \label{neighborhistogram}
\end{figure}


As shown in Figure \ref{neighborhistogram}, true neighbors of node $i$ occur most frequently in $\hat{\mathcal{L}}_{i}$. When ordering vertices in $\hat{\mathcal{L}}_{i}$ based on their frequency, most of the true neighbors appear at the top of the list. However, errors occur and false neighbors sometimes precede true neighbors. Note however that there are cases when the single neighborhood estimation performs much worse and omits many true neighbors. 

We have seen how histograms based on neighborhoods of pairs are useful in determining a likelihood ranking of possible neighbors of a given node. The top $t_i$ most frequent elements in $\hat{\mathcal{L}}_{i}$ are the most likely neighbors of $i$. The problem now becomes how to select this threshold value $t_i$ for each list. If $t_i$ is too small then true neighbors might be left out,  and if $t_i$ is too big then we will introduce false neighbors. 
We make an additional observation that improves on the accuracy of the above ordering obtained from $\hat{\mathcal{L}}_{i}$. Denote by $L$ the matrix formed from lists $\hat{\mathcal{L}}_{i}$ by letting $L_{ij} $ equal the frequency of node $ j $ in list $\hat{\mathcal{L}}_{i}$. To incorporate the symmetry between two neighboring nodes: if $i$ is a neighbor of $j$ then $j$ is also a neighbor of $i$, we build the symmetric matrix $ S = L + L^{T} $. The intuition here is to average out the votes received by nodes $i$ and $j$ in their respective rows. Suppose that $(i,j) \in E$, but $j$ does not rank highly in $\hat{\mathcal{L}}_{i}$. However, it may be the case that $i$ ranks highly in $\hat{\mathcal{L}}_{j}$, and helps in identifying that $(i,j)$ are indeed neighbors in $G$. One can think of this method as averaging out the bad information (noisy edges) and boosting up the good information (correct edges).

Finally, another alternative would be to first row normalize $L$ and then construct the symmetric matrix described above. We divide each row in $L$ by the largest entry in that row and obtain the row stochastic matrix $\bar{L}$, whose row $i$ may be thought of as ranked probabilities of the possible neighbors of $i$. We then build the row stochastic matrix $\bar{S} = \frac{1}{2}(\bar{L} + \bar{L}^{T})$, as described above.

As mentioned earlier, the main problem is finding the threshold $t_i$ for each row $i$ to separate the neighbors for non-neighbors of $i$. If we know a priori what the degree of each node is, then one way to pick the neighbors would be to select the most frequent $d_i$ entries in $\hat{\mathcal{L}}_{i}$. Alternatively, if we know that the graph is almost regular of degree $r$, or in other words that the average degree of the graph G is $r$ but the degree distribution has very little variance around $r$, then we can again select the top $r$ most frequent entries in $\hat{\mathcal{L}}_{i}$. Another way one can choose a threshold is to plot the frequency values in order and look for a big jump in the graph. This idea is illustrated in the right plot of Figure \ref{neighborhistogram}. Note how the frequency values decreases suddenly within two steps from 26 (for node 6) to 22 (for node 15) and then to 18 (for node 35). Such large sudden drops in the ordered list of frequency values hint at a good threshold point.

Table \ref{oraclecomparison} shows the results of an experiment that illustrates the above ideas when we have oracle information about the degree of each node. 
We observe that the symmetric matrix $S$ works better than just using $L$, and that $\bar{S}$ performs better than $S$. Note also that the type I and II errors are now more balanced both in the AND and OR case. When doing the estimation $\mathcal{N}^{1}$, in the AND case almost all the errors were coming from missing edges, while in the OR scenario almost all errors were given by false edges. Picking the threshold $t_i$ allows for a trade-off between these two type of errors. 

\begin{table}[!ht] 
\begin{center}
\begin{tabular}{|l||l|l||l|l|l|l|}
\hline
 &\multicolumn{3}{l|}{AND (Error type)} & \multicolumn{3}{l|}{OR (Error type)}\\
 \cline{2-7}
 Method      &  I &  II & Total    &  I &  II & Total \\
\hline\hline
$\mathcal{N}^{1}$ & 0.04   & 0.58 & 0.63 & 0.12 & 0.36 & 0.48\\
\hline
$\mathcal{N}^{2}$,$L$ & 0.04 & 0.27  & 0.31   &  0.32 & 0.09 & 0.41\\

\hline
$\mathcal{N}^{2}$,$S$ & 0.07 & 0.17  & 0.24   &  0.15 & 0.05 & 0.20\\

\hline
\rule{0cm}{0.4cm} $\mathcal{N}^{2}$,$\bar{S}$  & 0.05 & 0.14  & 0.19   &  0.135 & 0.045 & 0.18\\

\hline
\end{tabular}\end{center}
\caption{ A plot with the type I, type II, and total error probability for different ways of estimating the neighborhoods: $\mathcal{N}^{1}$ is obtained by the original method, and the remaining three rely on histograms of neighborhoods of pairs when then node degrees are given.}\label{comparevoting}
\label{oraclecomparison}
\end{table}

It would be interesting to compare the results of our new pair neighborhood recovery to the original single neighborhood method, when both techniques take into account the knowledge of node degree. One way to incorporate this information into the single neighborhood method is to avoid using 10-fold cross validation and replace it with the following procedure. We have seen that when regressing variable $i$ against all other variables, the elastic net method  does not return a single $\theta_i$ vector corresponding to the optimal $\lambda_1$ value, but rather computes an entire matrix (or sequence) where each row corresponds to a value of $\lambda_1$ at which an additional variable ''turns on''. Instead of picking the empirically optimal row with the 10-fold cross validation method as our $\theta_i$ vector, we can simply pick the first row which has $r_i$ nonzero entries, where $r_i$ is the degree of node $i$. One can also interpret the order in which the remaining $p-1$ variables ``turn on`` as a way of ranking the potential neighbors. A variable which ''turns on`` sooner on the elastic net path is more likely to be a neighbor of $i$ than a node which activates later on. This ranking can also be obtained by our pair-wise neighborhood union estimate, however we produce more than just a simple ranking. The frequency of each node in $\hat{\mathcal{L}}_i$ combines additional information, and one can interpret this ordering as a weighted ranking of possible neighbors of $i$. This additional information may capture longer range correlations that elude single neighborhood estimation, as suggested in a recent paper of Bento and Montanari [\ref{difficultgraph}].





\section{Conclusion}

In this paper we considered the problem of estimating the graph structure associated with a GMRF, the Ising model and the Potts model. Building on previous work using $l_1$ penalized neighborhood estimation, we experimented with the an additional $l_2$ penalty term.
Simulations across the three models show that a small but non-negligible $\lambda_2$ penalty term improves the edge recovery rates when the sample size is small by trading precision for recall. We make the observation that in the GMRF model, the addition of the $l_2$ penalty term does not have much influence on the recovery rates. Numerical simulations confirm our hypothesis that the lower bounds on the number of samples needed for recovery should not only be a function of the maximum degree $d$ and number of nodes $p$, but also of the edge density $\rho$ (or equivalently the average degree of the graph). We also introduce a new method for improving the neighborhood recovery by considering pair-wise neighborhood unions which produce a ranking of $p-1$ nodes in $G$ with respect to their likelihood of being adjacent to the remaining node. This can be thought of as a way to incorporate local information (rankings) at each node into a globally consistent edge structure estimation of the graph $G$.

\end{document}